\documentclass[aps,prl,twocolumn,amsfonts,amssymb,superscriptaddress]{revtex4} \usepackage{amsmath,mathrsfs} \usepackage{graphicx} \usepackage{hyperref}

\def\Tr{\operatorname{Tr}}

\def\>{\rangle}
\def\<{\langle}
\def\sH{\mathscr{H}}

\def\mE{\mathfrak{E}}

\def\rank{\operatorname{rank}}

\def\N#1{\left|\!\left|{#1}\right|\!
\right|}

\def\eps{\varepsilon}

\def\cqB{B_{\mathrm{cq}}^\eps}

\def\states{\mathfrak{S}}
\def\f{\mathsf{F}_{\mathrm{dil}}}
\def\mM{\mathcal{M}}

\newtheorem{theo}{Theorem}

\newtheorem{coro}{Corollary}
\newtheorem{lemma}{Lemma}
\newtheorem{definition}{Definition}

\renewcommand{\ge}{\geqslant}
\renewcommand{\le}{\leqslant}

\begin{document}

\title{Entanglement cost in practical scenarios}

\author{Francesco Buscemi} \email{buscemi@iar.nagoya-u.ac.jp}
\affiliation{Institute for Advanced Research, Nagoya University, Nagoya 464-8601, Japan}
\author{Nilanjana Datta} \email{n.datta@statslab.cam.ac.uk}
\affiliation{Statistical Laboratory, DPMMS, University of Cambridge,
  Cambridge CB3 0WB, UK}

\date{\today}

\begin{abstract}

  We quantify the one-shot entanglement cost of an arbitrary bipartite state, that is the minimum number of singlets needed by two distant parties to create a single copy of the state up to a finite accuracy, using local operations and classical communication only. This analysis, in contrast to the traditional one, pertains to scenarios of practical relevance, in which resources are finite and transformations can only be achieved approximately. Moreover, it unveils a fundamental relation between two well-known entanglement measures, namely, the Schmidt number and the entanglement of formation. Using this relation, we are able to recover the usual expression of the entanglement cost as a special case.

\end{abstract}

\maketitle



Among quantum information processing tasks, entanglement manipulation, namely, the interconversion between entangled states using only local transformations and classical communication, represents an important primitive. In this scenario, the abstract notion of entanglement becomes a fungible resource ``\emph{as real as energy}''~\cite{horo-rev}. This is one of the reasons for which intensive research has been devoted to the study of entanglement manipulations since the very early stages of Quantum Information Theory, making such an \emph{operational} theory of entanglement one of its biggest successes.

In this context, however, the word `operational' should not be confused with `practical'. Indeed, most results we have at present about entanglement resource theory rely on two unrealistic (and very strong) assumptions: 

\noindent
$(i)$ Many independent and identically distributed (i.i.d.) copies of the initial resource (e.g. the initial entangled state) are to be converted into many i.i.d. copies of the target state. This corresponds to assuming the absence of
correlations in the noisy (partially entangled) states which are either produced or consumed by the entanglement manipulation procedure;

\noindent
$(ii)$ The optimal interconversion rate is computed as the asymptotic input/output ratio, in the limit of infinitely many initial and final copies. 

\noindent
These two assumptions constitute what is usually called the asymptotic i.i.d. scenario. In order to establish a \emph{truly general} entanglement resource theory, then, one should drop both assumptions $(i)$ and $(ii)$. The highest possible degree of theoretical generality is described by the so-called \emph{one-shot} scenario, in which a single initial state has to be transformed into a single desired final state, up to a finite accuracy. Incidentally, this is indeed the scenario in which experiments are performed, since resources available in nature are typically
finite and correlated, and transformations can only be achieved approximately.

One end of such a generalized entanglement resource theory, namely, one-shot entanglement distillation, was considered by the present authors in~\cite{one-shot-distill}: there we described the case of two distant parties trying to convert, up to some fixed error $\eps$, a finite number of initially shared noisy bipartite entangled states into noiseless entanglement, i.e. singlets, using local operations and classical communication (LOCC) only. In this Letter we completely characterize the other end of the theory, namely, \emph{one-shot entanglement dilution}: here the goal is to utilize a finite amount of initial noiseless entanglement to produce (again, by LOCC and up to some fixed error $\eps$) a single bipartite target state $\rho_{AB}$, which might not be directly available otherwise. In this scenario, entanglement dilution is relevant as the `reverse' of entanglement distillation: it shows that singlets indeed provide a universal resource from which any bipartite state can be obtained by LOCC, quantifying, at the same time, the minimum amount of singlets needed (i.e. the \emph{cost}) to produce a given bipartite state.

Our main result~\cite{extramat} is a formula for the minimum number of singlets necessary for successfully producing a given target state $\rho_{AB}$ up to a finite error $\eps$. We refer to this quantity as the \emph{one-shot entanglement cost} $E_C^{(1)}(\rho_{AB};\eps)$. The formula we derive involves a generalized quantum relative entropy, namely, the relative R\'enyi entropy of order zero~\cite{ohya-petz}, and makes use of a smoothing procedure similar to that introduced in~\cite{renato}. When specialized to the asymptotic i.i.d. scenario, our formula yields the entanglement cost given in terms of the regularized entanglement of formation~\cite{bennett2,ent_cost}. This is in accordance with the claim that one-shot entanglement resource theory is more general than the asymptotic i.i.d. one. Finally, as a by-product of our findings, we are able to prove that two entanglement monotones, namely the entanglement of formation~\cite{bennett2} and the Schmidt number~\cite{terhal-pawel}, which were previously considered to be unrelated, are in fact directly connected, in the sense that the former is recovered from the latter by suitable smoothing and regularization, as explained below.

\emph{Basic concepts.}---In order to clearly state our main results, given in Theorems~\ref{thm_main} and~\ref{thm_main2} below, we first have to introduce some notations and definitions. Throughout the paper, the letter $\sH$ denotes finite dimensional Hilbert space, whereas $\states(\sH)$ denotes the set of states (or density operators, i.e. positive operators of unit trace) acting on $\sH$. Further, let $\openone$ denote the identity operator acting on $\sH$. Given a positive operator $\omega\ge 0$, we denote by $\Pi_\omega$ the projector onto its support, and, for a pure state $|\varphi\>$, we denote the projector $|\varphi\>\<\varphi|$ simply as $\varphi$. Moreover, given two Hilbert spaces $\sH_A$ and $\sH_B$, of dimensions $d_A$ and $d_B$ respectively, with two given orthonormal bases $\{|i_A\rangle\}_{i=1}^{d_A}$ and $\{|i_B\rangle\}_{i=1}^{d_B}$, we define the canonical maximally entangled state (MES) in $\sH_A\otimes\sH_B$ of Schmidt number $M \le\min \{d_A,d_B\}$ to be $|\Psi^+_M\>= M^{-1/2} \sum_{i=1}^M |i_A\rangle\otimes |i_B\rangle$.

Information-theoretical protocols, since Shannon, are usually characterized in term of suitable entropic quantities. In Quantum Information Theory too, entropic quantities like the von Neumann entropy, the conditional entropy, and the mutual information are often encountered. All these quantities can in fact be derived from the \emph{quantum relative entropy}~\cite{ohya-petz}, which is defined, for a state $\rho$ and an operator $\sigma\ge 0$, as \begin{equation} \nonumber S_r(\rho\|\sigma):=\left\{ \begin{split}
      &\Tr[\rho\log\rho -\rho\log\sigma],\textrm{ if }\Pi_\rho\le\Pi_\sigma,\\
      &+\infty,\textrm{ otherwise}.  \end{split} \right.  \end{equation} (The logarithm in the above equation and in what follows is taken to base 2.) For example, the von Neumann entropy of a state $\rho$, defined as $S(\rho):=-\Tr[\rho\log\rho]$, can be equivalently written as $S(\rho)=-S_r(\rho\|\openone)$. Our main results are however expressed in terms of an alternative relative entropy, namely, the \emph{relative R\'enyi entropy of order zero}, which, for a state $\rho$ and an operator $\sigma\ge 0$, is defined as \begin{equation}\nonumber S_0(\rho\|\sigma):=\left\{ \begin{split}
      &-\log\Tr[\Pi_\rho\ \sigma],\textrm{ if }\Tr[\Pi_\rho\Pi_\sigma]\neq 0,\\
      &+\infty,\textrm{ otherwise}.  \end{split}\right.  \end{equation} From these two relative entropies, $S_r$ and $S_0$, we define the corresponding conditional entropy of a given bipartite state $\rho_{AB}$ given a state $\sigma_B$ as \begin{equation}
  \label{eq:19}
  H_{\star}(\rho_{AB}|\sigma_B):=-S_{\star}(\rho_{AB}\|\openone_A\otimes\sigma_B),
\end{equation}
and the conditional entropy of $\rho_{AB}$ given the subsystem $B$ as
\begin{equation}
  \label{eq:22}
  H_{\star}(\rho_{AB}|B):=\max_{\sigma_B\in\states(\sH_B)}H_{\star}(\rho_{AB}|\sigma_B),
\end{equation}
for $\star\in\{r,0\}$. It turns out (see e.g. Lemma~6 in~\cite{q1}) that $H_r(\rho_{AB}|B)=H_r(\rho_{AB}|\rho_B)=S(\rho_{AB})-S(\rho_B)$, where $\rho_B=\Tr_A[\rho_{AB}]$, for any given $\rho_{AB}$. However, in general, $H_0(\rho_{AB}|B)\neq H_0(\rho_{AB}|\rho_B)$.

It is also convenient to introduce, for any given decomposition of a bipartite state
$\rho_{AB}$ into a pure-state ensemble $\mE=\{p_i,|\phi^i_{AB}\>\}$ such that $\sum_ip_i\phi^i_{AB}=\rho_{AB}$, the tripartite classical-quantum (c-q) state
\begin{equation}
  \label{eq:28}
  \rho_{RAB}^\mE:=\sum_ip_i|i\>\<i|_R\otimes\phi_{AB}^i,
\end{equation}
where $R$ denotes an auxiliary classical system represented by the fixed orthonormal basis $\{|i_R\>\}$. Given a pure-state ensemble $\mE$, let $\rho_A^i:=\Tr_B[\phi_{AB}^i]$, for all $i$. 

As noted earlier, in the realistic scenario of finite entanglement resources and imperfect transformations, one is compelled to allow for a non-vanishing error, say $\eps$, in achieving the final desired state. This error $\eps$ manifests itself as a ``smoothing'' of the underlying information-theoretical quantity characterizing the task, which in our case turns out to be a conditional R\'enyi entropy of order zero. This fact leads us to define, in analogy with~\cite{renato}, a smoothing as follows: for any $\eps\ge0$ and any pure-state ensemble $\mE=\{p_i,|\phi^i_{AB}\>\}$ of $\rho_{AB}$, we define the \emph{c-q--smoothed conditional zero-R\'enyi entropy} of the c-q state $\rho_{RA}^\mE:=\Tr_B[\rho_{RAB}^\mE]=\sum_i p_i|i\>\<i|_R\otimes \rho_A^i$, given $R$, as \begin{equation}\label{eq:2} H_0^\eps(\rho_{RA}^\mE|R):=\min_{\omega_{RA}\in\cqB(\rho_{RA}^\mE)}H_0(\omega_{RA}|R), \end{equation} where the minimum is taken over classical-quantum operators belonging to the set $\cqB(\rho_{RA}^\mE)$ defined, for any pure-state ensemble $\mE=\{p_i,|\phi^i_{AB}\>\}$ of $\rho_{AB}$, as follows: \begin{equation}
  \nonumber
  \cqB(\rho_{RA}^\mE):=\left\{\omega_{RA}\ge
    0 \left|
      \begin{split}
        &\omega_{RA}=\sum_i|i\>\<i|_R\otimes\omega^i_A\\
        &\&\ \N{\omega_{RA}-\rho_{RA}^\mE}_1\le\eps
      \end{split}
    \right.\right\},
\end{equation}
with $\N{X}_1:=\Tr|X|$. The basis $\{|i_R\>\}$ used in the above definition is the same as that appearing in eq.~(\ref{eq:28}). Note that operators in $\cqB(\rho_{RA}^\mE)$ are actually very close to being density operators, since $1-\eps\le\Tr[\omega_{RA}]\le 1+\eps$, for any $\omega_{RA}\in \cqB(\rho_{RA}^\mE)$.\smallskip

\emph{Main result.}---Two parties, Alice and Bob, share a {\em{single copy}} of a maximally entangled state $|\Psi^+_M\>$ of Schmidt number $M$, and wish to convert it into a given bipartite target state $\rho_{AB}$ using an LOCC map $\Lambda$. We refer to the protocol used for this conversion as {\em{one-shot entanglement dilution}}. For sake of generality, we consider the situation where the final state of the protocol is $\eps$-close to the target state with respect to a suitable distance measure, for any given $\eps\ge 0$. As a measure of closeness, we choose here the (squared) fidelity, which is defined, for states $\rho$ and $\sigma$, as $F^2(\rho,\sigma):=\left(\Tr|\sqrt{\rho}\sqrt{\sigma}|\right)^2$. In this way, defining the fidelity of the protocol to be $F^2(\Lambda(\Psi_M^+),\rho_{AB})$, we require $F^2(\Lambda(\Psi_M^+),\rho_{AB})\ge 1-\eps$. Further, for any given initial resource $|\Psi^+_M\>$ and any given target state $\rho_{AB}$, we denote the optimal fidelity of one-shot entanglement dilution as
\begin{equation}
  \nonumber
 \f(\rho_{AB},M):=\max_{\Lambda\in\mathrm{LOCC}}F^2(\Lambda(\Psi_M^+),\rho_{AB}).
\end{equation}


\begin{definition}[One-shot entanglement cost] For any given $\rho_{AB}$ and $\eps\ge0$, the \emph{one-shot entanglement cost} is defined as follows: 
\begin{equation}\nonumber
E_C^{(1)}(\rho_{AB};\eps):=\min_{M\in\mathbb{N}}\left\{\log M: \f(\rho_{AB},M)\ge 1-\eps\right\}.
\end{equation}\label{def:ent-cost}
\end{definition}

Notice that, by its very definition, the one-shot entanglement cost $E_C^{(1)}(\rho_{AB};\eps)$ constitutes, for any $\eps\ge 0$, an entanglement (weak) monotone, in that it cannot increase under the action of an LOCC map,~\cite{footnote1}. As mentioned earlier, the smoothing here emerges naturally from a purely operational consideration, in the sense that it is a natural consequence of the finite accuracy we allow in the protocol. This is in contrast to the approach adopted in Ref.~\cite{mora}, where a smoothing is instead introduced axiomatically.

Our main result is given by the following theorem:

\begin{theo}\label{thm_main} For any given target state $\rho_{AB}$ and any given error parameter $\eps\ge 0$, the one-shot entanglement cost under LOCC, corresponding to an error less than or equal to $\eps$, satisfies the following bounds: \begin{equation}
  \nonumber
  \min_{\mE}H_0^{2\sqrt{\eps}}(\rho_{RA}^\mE|R)\le
  E_C^{(1)}(\rho_{AB};\eps)
 \le\min_{\mE}H_0^{\eps/2}(\rho_{RA}^\mE|R),
\end{equation}
where the minimum is taken over all pure-state ensemble decompositions $\mE=\{p_i,|\phi^i_{AB}\>\}$ of $\rho_{AB}$, and $\rho_{RA}^\mE=\Tr_B[\rho_{RAB}^\mE]$, with $\rho_{RAB}^\mE$ being the tripartite extension of $\rho_{AB}$ defined in~(\ref{eq:28}).
\end{theo}

For any given $\eps\ge 0$, Theorem~\ref{thm_main} essentially identifies $\min_{\mE}H_0^{\eps}(\rho_{RA}^\mE|R)$ as the quantity representing the one-shot entanglement cost $ E_C^{(1)}(\rho_{AB};\eps)$,~\cite{footnote2}.

The theory developed here not only provides a complete characterization of the one-shot entanglement cost, it also yields a simple proof of a fundamental asymptotic result.  It is known~\cite{ent_cost} that the asymptotic entanglement cost $E_C(\rho_{AB})$ of preparing a bipartite state $\rho_{AB}$ is equal to the \emph{regularized entanglement of formation}, defined as, \begin{equation}
  \label{eq:35}
  E_F^\infty(\rho_{AB}):=\lim_{n\to\infty}\frac 1nE_F(\rho_{AB}^{\otimes n}),
\end{equation}
where $E_F(\rho_{AB}):=\min_\mE\sum_ip_iS(\rho_A^i)$ denotes the entanglement of formation of the state $\rho_{AB}$ \cite{bennett2}. Applying our main result, Theorem~\ref{thm_main}, to the case of multiple ($n$) copies of the bipartite state $\rho_{AB}$, and taking the asymptotic limit ($n\to\infty$) yields a new proof of the identity $E_C(\rho_{AB})=E_F^\infty(\rho_{AB})$:

\begin{theo}\label{thm_main2}
For any given target state $\rho_{AB}$, the following identity holds:
\begin{equation}\label{eq:thm_main2}
\lim_{\eps\to 0^+}\lim_{n\to\infty}\frac 1n E_C^{(1)}(\rho_{AB}^{\otimes n};\eps)=E_F^\infty(\rho_{AB}).
\end{equation}
\end{theo}

Theorem~\ref{thm_main2}, together with the results in Ref.~\cite{ent_cost}, establishes that the asymptotic entanglement cost is alternatively expressible as the regularized one-shot entanglement cost, in the limit $\eps\to 0^+$.

The theorems stated above emphasize the generality and two-fold relevance of the one-shot analysis: on one hand, it gives a complete description of realistic scenarios of entanglement dilution, on the other hand, it provides a unified theoretical framework from which previous results can be derived as special cases.

\emph{Discussion.}---In the case of perfect (zero-error) entanglement dilution, corresponding to the case $\eps=0$, Theorem~\ref{thm_main} says that the corresponding one-shot entanglement cost is given by
\begin{equation}\label{eq:zero-err-cost}
  E_C^{(1)}(\rho_{AB};0)=\min_{\mE} H_0(\rho_{RA}^\mE|R).
\end{equation}
The above equation can be made more explicit as follows:
\begin{equation}\nonumber
  E_C^{(1)}(\rho_{AB};0)=\min_\mE\max_i\log\Tr\left[\Pi_{\rho^i_A}\right],
\end{equation}
where, for any given pure-state ensemble decomposition $\mE=\{p_i,|\phi^i_{AB}\>\}$ of $\rho_{AB}$, $\rho^i_A:=\Tr_B[\phi^i_{AB}]$. The quantity on the right-hand side of the equation above coincides with the logarithm of the Schmidt number (log-Schmidt number, for short) of the mixed state $\rho_{AB}$, introduced and studied in~\cite{terhal-pawel}. In~\cite{hayashi}, the same quantity, was denoted as $E_{sr}(\rho_{AB})$, and was shown to characterize the zero-error entanglement cost $E_C^{(1)}(\rho_{AB};0)$. However, until now, there was a gap in the theory of entanglement dilution, in the sense that it was unclear how these zero-error results could be related to the usual notion of entanglement cost, for which the error vanishes only in the asymptotic limit.

The results we presented above show that it is indeed possible to fill such a gap by suitably smoothing the zero-error quantities. In fact, let us introduce a smoothed log-Schmidt number as follows:
\begin{equation}\label{eq:esrsm}
E_{sr}^\eps(\rho_{AB}):=\min_{\omega_{AB}\in C_\eps(\rho_{AB})}E_{sr}(\omega_{AB}), 
\end{equation}
where now the smoothing is performed with respect to the compact set of normalized states $C_\eps(\rho_{AB})$ centered at $\rho_{AB}$ defined as: \begin{equation}\nonumber \begin{split}
    &C_\eps(\rho_{AB})\\ &:=\left\{\omega_{AB}\in\states(\sH_A\otimes\sH_B)\left|F^2(\omega_{AB},\rho_{AB})\ge1-\eps\right.\right\}.
\end{split}
\end{equation}
Then, using the arguments given below, one can prove that, for any $\eps\ge0$, the identity $E_C^{(1)}(\rho_{AB};\eps)=E_{sr}^\eps(\rho_{AB})$ holds. First, for any $\omega_{AB}\in C_\eps(\rho_{AB})$, $E_{sr}(\omega_{AB})$ singlets can be used to create, with zero-error, the state $\omega_{AB}$, which is, by construction, $\eps$-close to $\rho_{AB}$. This proves that $E_C^{(1)}(\rho_{AB};\eps) \le E_{sr}^\eps(\rho_{AB})$. For the other direction, let us assume that $E_C^{(1)}(\rho_{AB};\eps)<E_{sr}^\eps(\rho_{AB})$. Definition~\ref{def:ent-cost} then implies that, with $E_C^{(1)}(\rho_{AB};\eps)$ singlets, it is possible to create a state, say $\tilde\omega_{AB}$, which is $\eps$-close to $\rho_{AB}$. This in turn implies that $\tilde\omega_{AB}\in C_\eps(\rho_{AB})$, with $E_{sr}(\tilde\omega_{AB})=E_C^{(1)}(\rho_{AB};\eps)<E_{sr}^\eps(\rho_{AB})$, which contradicts the fact that $E_{sr}^\eps(\rho_{AB})$ is defined as a minimum in~(\ref{eq:esrsm}).

We hence obtain the following corollary of Theorem~\ref{thm_main2}:

\begin{coro} For any given state $\rho_{AB}$, the entanglement of formation $E_F(\rho_{AB})$ and the log-Schmidt number $E_{sr}(\rho_{AB})$ are related as follows: \begin{equation}\label{eq:corol}
\lim_{\eps\to 0^+}\lim_{n\to\infty}\frac 1n E_{sr}^\eps(\rho_{AB}^{\otimes n})=E_F^\infty(\rho_{AB}).
\end{equation}
\end{coro}


\emph{Essence of proofs.}---We present here only the main steps of the proofs of the results stated above. The interested reader is referred to~\cite{extramat} for detailed derivations. The proof of Theorem~\ref{thm_main} relies on the following lemma:
\begin{lemma}[\cite{bowen-datta,hayashi}]\label{fidelity}
For any given bipartite state $\rho_{AB}$, the optimal dilution fidelity is given by \begin{equation}
  \label{eq:8}
  \f(\rho_{AB},M)=\max_\mE\sum_ip_i\sum_{j=1}^M\lambda_j^{(i)},
\end{equation}
where the maximum is over all pure-state decomposition $\mE=\{p_i,|\phi_{AB}^i\>\}$ of $\rho_{AB}$, and $\{\lambda_j^{(i)}\}_j$ are the eigenvalues of $\rho^i_A=\Tr_B[\phi_{AB}^i]$, arranged in non-increasing order.
\end{lemma}
Using this lemma and Definition~\ref{def:ent-cost}, we can prove that $E_C^{(1)}(\rho_{AB};\eps)=\min_{\mE}E^\eps(\mE)$, where
\begin{equation}\nonumber
  \begin{split}
    &E^\eps(\mE)\\    &:=\min_{\{\Pi^i_A\}}\left\{\max_i\log\Tr\left[\Pi^i_A\right]\left|\sum_ip_i\Tr\left[\Pi^i_A\rho^i_A\right]\ge1-\eps\right.\right\},
  \end{split}
\end{equation}
where $\{\Pi^i_A\}$ is an unconstrained set of projectors, that is, not necessarily orthogonal nor complete. The proof of Theorem~\ref{thm_main} then reduces to proving that $H_0^{2\sqrt{\eps}}(\rho_{RA}^\mE|R) \le E^\eps(\mE)\le H_0^{\eps/2}(\rho_{RA}^\mE|R)$, for any ensemble $\mE$ and any $\eps\ge 0$. This is done by standard tools like convexity arguments and the ``gentle measurement'' lemma~\cite{gentle}.

As regards the asymptotic result of Theorem~\ref{thm_main2}, the starting point is to note that the entanglement of formation itself can be expressed as a conditional entropy $E_F(\rho_{AB})=\min_{\mE}H_r(\rho_{RA}^\mE|R)$, in close analogy with the expression~(\ref{eq:zero-err-cost}) of the zero-error one-shot entanglement cost. Theorem~\ref{thm_main2} then reduces to the identity
\begin{equation}\label{eq:asymp2}
\begin{split}
\lim_{\eps\to 0^+}&\lim_{n\to\infty}\frac 1n \min_{\mE_n}H_0^{\eps}(\rho_{R_nA_n}^{\mE_n}|R_n)\\
=&\lim_{n\to\infty}\frac 1n \min_{\mE_n}H_r(\rho_{R_nA_n}^{\mE_n}|R_n) \equiv E_F^\infty(\rho_{AB}),
\end{split}
\end{equation}
where $\mE_n$ denotes a pure-state ensemble decomposition $\{p_i^n,|\phi^i_{A_nB_n}\>\}$ of $\rho_{AB}^{\otimes n}$, such that $\rho_{AB}^{\otimes n}=\sum_ip_i^n \phi^i_{A_nB_n}$, and $\rho_{R_nA_n}^{\mE_n}=\Tr_{\sH_B^{\otimes n}}[\rho^{\mE_n}_{R_nA_nB_n}]$, with $\rho_{R_nA_nB_n}^{\mE_n}$ denoting the c-q extension of $\rho_{AB}^{\otimes n}$ as in equation~(\ref{eq:28}). The identity~(\ref{eq:asymp2}) is proved by employing the information spectrum method~\cite{info-spect}, results of~\cite{nila}, and a generalized version of Stein's lemma established in~\cite{brandao-plenio}.

\emph{Acknowledgments}---FB acknowledges support from Japan's MEXT-SCF Program for Improvement of Research Environment for Young Researchers. ND acknowledges support from the European Community's Seventh Framework Programme (FP7/2007-2013) under grant agreement number 213681. This work was completed when FB was visiting the Statistical Laboratory of the University of Cambridge.

\newpage

\section{Detailed proof of Theorem~\ref{thm_main}}

For a given bipartite state $\rho_{AB}$, let $\mE=\{p_i,|\phi^i_{AB}\>\}$ be an ensemble of pure states such that $\sum_ip_i\phi^i_{AB}=\rho_{AB}$. We introduce the following quantity:
\begin{equation}
  \label{eq:1}
\begin{split}
&E^\eps(\mE)\\
&:=\min_{\{\Pi^i_A\}}\left\{\max_i\log\Tr\left[\Pi^i_A\right]\left|\sum_ip_i\Tr\left[\Pi^i_A\rho^i_A\right]\ge1-\eps\right.\right\},
\end{split}
\end{equation}
where $\rho^i_A:=\Tr_B[\phi^i_{AB}]$, and $\{\Pi^i_A\}$ is an
unconstrained set of projectors, that is, not necessarily orthogonal nor complete.

We first prove the following lemma relating $E^\eps(\mE)$ to the c-q--smoothed conditional zero-R\'enyi entropy appearing in the statement of Theorem~\ref{thm_main}:

\begin{lemma}\label{lemma:useful}
For any $\eps\ge 0$, and any choice of the pure-state ensemble $\mE$ for
$\rho_{AB}$, the following holds:
\begin{equation}
  \label{eq:4}
 H_0^{2\sqrt{\eps}}(\rho_{RA}^\mE|R)\le E^\eps(\mE)\le H_0^{\eps/2}(\rho_{RA}^\mE|R),
\end{equation}
where $\rho_{RA}^\mE$ is the reduced state obtained from the tripartite extension $\rho_{RAB}^\mE$ defined in~(\ref{eq:28}). $\square$
\end{lemma}

\noindent{\bf Proof}. We first prove the bound
\begin{equation}
  \label{eq:17}
H_0^{\eps}(\rho_{RA}^\mE|R)\ge E^{2\eps}(\mE).
\end{equation}
Let $\overline\omega_{RA}=\sum_i|i\>\<i|_R\otimes\overline\omega_A^i$ in
$\cqB(\rho_{RA}^\mE)$ be the operator achieving the minimum
in~(\ref{eq:2}). The projection onto its support is given by
$\Pi_{{\overline\omega}_{RA}}=\sum_i|i\>\<i|_R\otimes\Pi_{\overline\omega_A^i}$. Hence, $\overline\omega_{RA}\in\cqB(\rho_{RA}^\mE)$ yields a set of projectors
$\{\Pi_{\overline\omega^i_A}\}$ for which
\begin{equation}
  \nonumber
\begin{split}
  &\sum_ip_i\Tr[\Pi_{\overline\omega^i_A}\rho_A^i]=\Tr[\Pi_{\overline\omega_{RA}}\rho_{RA}^\mE]\\
  =&\Tr[\Pi_{\overline\omega_{RA}}\overline\omega_{RA}]+\Tr[\Pi_{\overline\omega_{RA}}(\rho_{RA}^\mE-\overline\omega_{RA})]\\
  \ge&1-\eps-\eps=1-2\eps.
\end{split}
\end{equation}
In the last line we have made use of the fact that $\overline\omega_{RA}\in\cqB(\rho^\mE_{RA})$, due to which $\Tr[\overline\omega_{RA}]\ge 1-\eps$. This implies that the set of projectors $\{\Pi_{\overline\omega^i_A}\}$ satisfies the condition required in definition~(\ref{eq:1}) of $E^{2\eps}(\mE)$, hence proving~(\ref{eq:17}).

We now prove the lower bound
\begin{equation}
  \label{eq:21}
  E^{\eps}(\mE)\ge  H_0^{2\sqrt{\eps}}(\rho_{RA}^\mE|R).
\end{equation}
Let $\left\{\overline\Pi^i_A\right\}$ be the set of projectors achieving the minimum in eq.~(\ref{eq:1}). Therefore, $\sum_ip_i\Tr\left[\overline\Pi^i_A\rho^i_A\right]\ge 1-\eps$. For later convenience, let us set $\eps_i:=1-\Tr\left[\overline\Pi^i_A\rho^i_A\right]$, so that $\sum_ip_i\eps_i\le\eps$. Let us define $\overline\omega_A^i:=\overline\Pi^i_A\rho_A^i\overline\Pi^i_A$ and $\overline\omega_{RA}:=\sum_ip_i|i\>\<i|_R\otimes \overline\omega_A^i$.
The so-called Gentle Measurement Lemma~\cite{gentle} guarantees that $\N{\overline\omega_A^i-\rho_A^i}\le2\sqrt{\eps_i}$, for all $i$. Also, by the concavity of $x\mapsto\sqrt{x}$, we have:
\begin{equation}
  \label{eq:23}
\begin{split}
  \N{\overline\omega_{RA}-\rho_{RA}^\mE}_1&=\sum_ip_i\N{\overline\omega_A^i-\rho_A^i}_1\\
  &\le\sum_ip_i2\sqrt{\eps_i}\\
  &\le2\sqrt{\sum_ip_i\eps_i}\le2\sqrt{\eps}.
\end{split}
\end{equation}
The above inequalities prove that $\overline\omega_{RA}\in
B_{\mathrm{cq}}^{2\sqrt{\eps}}(\rho_{RA}^\mE)$. Moreover, since, by
construction, $\Pi_{\overline\omega^i_A}\le\overline\Pi^i_A$ for all $i$,
we obtain eq.~(\ref{eq:21}). $\blacksquare$\bigskip

With Lemma~\ref{lemma:useful} in hand, the proof of Theorem~\ref{thm_main} reduces to proving the following identity:
\begin{equation}
  \label{eq:30}
  E_C^{(1)}(\rho_{AB};\eps)=\min_{\mE}E^\eps(\mE).
\end{equation}
We split the proof of this identity into  Lemma~\ref{lemma:3} and 
Lemma~\ref{lemma:4} below.

\begin{lemma}[Direct part]\label{lemma:3} For any $\eps\ge 0$,
  \begin{equation}
    \nonumber
    E_C^{(1)}(\rho_{AB};\eps)\le\min_{\mE}E^\eps(\mE).
  \end{equation}
\end{lemma}

\noindent{\bf Proof}. From Lemma~\ref{fidelity},
\begin{equation}
  \label{eq:9}
  \f(\rho_{AB},M)=\max_\mE\sum_ip_i\Tr[Q_M^i\ \rho^i_A],
\end{equation}
where, for each $i$, $Q_M^i$ is the projector onto the eigenvectors associated with the $M$ largest eigenvalues of $\rho^i_A$. Let us now fix an ensemble
decomposition $\overline\mE:=\left\{\overline p_i,|\overline \phi^i_{AB}\>\right\}$ for $\rho_{AB}$, and choose the integer $\overline M$ such that $\log\overline M=E^\eps\left(\overline\mE\right)$. Then, from definition~(\ref{eq:1}), we know that there 
exists a set of projectors $\{\Pi^i_A\}$, with $\rank
\Pi^i_A\le\overline M$ for all $i$, such that $\sum_i\overline p_i\Tr[\Pi^i_A\overline
\rho^i_A]\ge1-\eps$. This implies that $\log\overline M$ is an
$\eps$-achievable rate, since
\begin{equation}
  \label{eq:10}
\begin{split}
  \f(\rho_{AB},\overline M)&\ge\sum_i\bar p_i\Tr\left[\overline Q_{\overline M}^i\overline\rho_A^i\right]\\
&\ge
  \sum_i\bar p_i\Tr\left[\Pi^i_A\overline
\rho^i_A\right]\ge1-\eps,
\end{split}
\end{equation}
where $\overline Q_{\bar M}^i$ is, for each $i$, the projector onto the $\overline M$ largest
eigenvalues of $\overline\rho_A^i$. The second inequality in~(\ref{eq:10})
is due to the fact that, for any projector $\Pi^i_A$ with
$\rank\Pi^i_A\le\overline M$, $\Tr[\Pi^i_A\overline\rho_A^i]\le\Tr\left[\overline Q_{\overline M}^i\overline\rho_A^i\right]$. Hence $E^\eps\left(\overline\mE\right)$ is itself an $\eps$-achievable rate for any choice of $\overline\mE$, and the statement of the lemma follows. $\blacksquare$\bigskip

\begin{lemma}[Weak converse]\label{lemma:4} For any $\eps\ge 0$,
  \begin{equation}
    \label{eq:6}
    E_C^{(1)}(\rho_{AB};\eps)\ge\min_{\mE}E^\eps(\mE).
  \end{equation}
\end{lemma}

\noindent{\bf Proof}. Let $\log M$ be an $\eps$-achievable rate. This
is equivalent to saying that $\f(\rho_{AB},M)\ge1-\eps$. In the
following, we prove that this implies that
\begin{equation}
  \label{eq:11}
  \log M\ge\min_\mE
E^\eps(\mE).
\end{equation}
Let $\overline\mE:=\{p_i,|\phi^i_{AB}\>\}$ be the ensemble decomposition of
$\rho_{AB}$ achieving $\f(\rho_{AB},M)$ in~(\ref{eq:8}), and consider
the Schmidt decomposition of its elements $|\phi^i_{AB}\>
=\sum_j\sqrt{\lambda_j^{(i)}} |j_A^{(i)}\>|j_B^{(i)}\>$, where the Schmidt
coefficients $\{\lambda^{(i)}_j\}_j$ are arranged in non-increasing order for all $i$. The optimal dilution fidelity given by eq.~(\ref{eq:8}) can then be expressed as
\begin{equation}
  \label{eq:12}
  \f(\rho_{AB},M)=\sum_ip_i\Tr[\overline\omega^i_A]\ge1-\eps,
\end{equation}
where
\begin{equation}
  \label{eq:13}
  \overline\omega^i_A:=\sum_{j=1}^M\lambda_j^{(i)}|j^{(i)}\>\<j^{(i)}|_A.
\end{equation}
We now proceed by observing that
\begin{equation}
  \label{eq:14}
\begin{split}
  &E^\eps\left(\overline\mE\right)\\
  &\le\min_{\{\omega^i_A\}}\left\{\max_i\log\Tr\left[\Pi_{\omega^i_A}\right]\left| \begin{split}
        &\sum_ip_i\Tr[\omega^i_A]\ge1-\eps\\
        &\&\ \omega^i_A\le\rho^i_A,\ \forall i\\
      \end{split} \right.\right\}.
\end{split}
\end{equation}
This is due to the fact that, for any set of operators $\{\omega^i_A\}$ satisfying both conditions at the right hand side, the corresponding set of projectors $\{\Pi_{\omega^i_A}\}$ satisfies the conditions required in the definition~(\ref{eq:1}) of $E^\eps\left(\overline\mE\right)$. This is because $1-\eps\le \sum_ip_i\Tr\left[\Pi_{\omega^i_A}\omega^i_A\right]\le \sum_ip_i\Tr\left[\Pi_{\omega^i_A}\rho^i_A\right]$. In particular, also the set of subnormalized density operators $\{\overline\omega^i_A\}$ defined by~(\ref{eq:13}) satisfies both conditions required at the right hand side of eq.~(\ref{eq:14}), since $\overline\omega^i_A\le\rho^i_A$ for all $i$ (by definition), and $\sum_ip_i\Tr[\overline\omega^i_A] \ge 1-\eps$ (by eq.~(\ref{eq:12})). We then have:
\begin{equation}
  \label{eq:15}
\begin{split}
\min_\mE E^\eps(\mE) \le E^\eps(\bar\mE)&\le
\max_i\log\Tr\left[\Pi_{\bar\omega^i_A}\right]\\
&\le\log M,
\end{split}
\end{equation}
for any $\eps$-achievable rate $\log M$. $\blacksquare$

\section{Detailed proof of Theorem~\ref{thm_main2}}
\label{asymp}

By defining the asymptotic entanglement cost as
\begin{equation}\label{eq:asym-cost}
  E_C(\rho_{AB}):=\lim_{\eps\to 0^+}\lim_{n\to\infty}\frac 1n E_C^{(1)}(\rho_{AB}^{\otimes n};\eps),
\end{equation}
we prove that
\begin{equation}
  E_C(\rho_{AB})=E_F^\infty(\rho_{AB}).
\end{equation}
Hence, we also prove indirectly that definition~(\ref{eq:asym-cost}) is equivalent to the alternative definitions of asymptotic entanglement cost proposed in Ref.~\cite{ent_cost}. We split the proof of Theorem~\ref{thm_main2} into Lemma~\ref{lem:asym-conv} and Lemma~\ref{lem:asym-dir} below. In the following, $\mE_n$ denotes a pure-state ensemble decomposition $\{p_i^n,|\phi^i_{A_nB_n}\>\}$ of $\rho_{AB}^{\otimes n}$, such that $\rho_{AB}^{\otimes n}=\sum_ip_i^n \phi^i_{A_nB_n}$. Notice that, even though the state $\rho_{AB}^{\otimes n}$ is in product form, the pure states $\{|\phi^i_{A_nB_n}\>\}$ in some of its decompositions may well be entangled. For the reader's convenience, we recall that the entanglement of formation $E_F(\rho_{AB}):=\min_{\mE}\sum_iS(\rho^i_A)$ can itself be written as a conditional entropy: $E_F(\rho_{AB})=\min_{\mE}H_r(\rho_{RA}^\mE|R)= \min_{\mE}H_r(\rho_{RA}^\mE|\rho_{R}^\mE)$, where $\rho_{RA}^\mE= \Tr_B \rho_{RAB}^\mE$, with $\rho_{RAB}^\mE$ being the tripartite extension of the state $\rho_{AB}$, defined by~(\ref{eq:28}). Hence, $E_F^\infty(\rho_{AB})=\lim_{n\to\infty}\frac 1n \min_{\mE_n}H_r(\rho^{\mE_n}_{R_nA_n}|\rho_{R_n}^{\mE_n})$, where $\rho^{\mE_n}_{R_nA_n}$ is as in 
(11).

\begin{lemma}\label{lem:asym-conv}
The following holds:
\begin{equation}\label{eq:final}
E_C(\rho_{AB})\ge E_F^\infty(\rho_{AB}).\ \square
  \end{equation}
\end{lemma}

\noindent{\bf Proof}. We start with the lower bound in Theorem~\ref{thm_main}. For any $\eps\ge0$ and any $n\in\mathbb{N}$, this gives
\begin{equation}
\begin{split}
&\frac 1n E_C^{(1)}\left(\rho_{AB}^{\otimes n};\frac{\eps^2}4\right)\\
\ge&\frac 1n\min_{\mE_n}H_0^\eps\left(\rho^{\mE_n}_{R_nA_n}|R_n\right)\\
=&\frac 1n\min_{\mE_n}\min_{\omega_{R_nA_n}^n\in\cqB\left(\rho_{R_nA_n}^{\mE_n}\right)}H_0(\omega^n_{R_nA_n}|R_n)\\
=&\frac 1nH_0(\overline\omega^n_{R_nA_n}|R_n)\\
\ge& \frac 1nH_r(\overline\omega^n_{R_nA_n}|R_n)\\
=&\frac 1nH_r(\overline\omega^n_{R_nA_n}|\overline\omega_{R_n}^n)\\
\ge&\frac 1n\min_{\mE_n}H_r\left(\rho^{\mE_n}_{R_nA_n}\left|\rho_{R_n}^{\mE_n}\right.\right)-O(\eps)-O(1/n),
\end{split}  
\end{equation}
\noindent where: in the fourth line, $\overline\omega^n_{RA}$ is the minimizing operator for the minimizing pure-state ensemble $\overline\mE_n$ of $\rho_{AB}^{\otimes n}$; in the fifth line we use the fact that $H_0(\overline\omega^n_{R_nA_n}|R_n)\ge H_r(\overline\omega^n_{R_nA_n}|R_n)$ which follows from the well-known fact that $S_0(\rho\|\sigma)\le S_r(\rho\|\sigma)$~\cite{ohya-petz}; the sixth line follows from Lemma~6 in~\cite{q1}.  The last approximation comes by applying Fannes' inequality to $H_r(\overline\omega^n_{R_nA_n}|\overline\omega_{R_n}^n) =S(\overline\omega^n_{R_nA_n})-S(\overline\omega_{R_n}^n)$. Then, by considering the limit $n\to\infty$ followed by $\eps\to 0^+$, we arrive at eq.~(\ref{eq:final}). $\blacksquare$\bigskip

\begin{lemma}\label{lem:asym-dir}
The following holds:
\begin{equation}\label{eq:to_prove}
 E_C(\rho_{AB})\le E_F^\infty(\rho_{AB}).\ \square
\end{equation}
\end{lemma}

\noindent{\bf Proof}. From the upper bound in Theorem~\ref{thm_main} we obtain
\begin{equation}\label{qwerty}
\frac 1n  E_C^{(1)}(\rho_{AB}^{\otimes n};2\eps)\le\frac 1n\min_{\mE_n}H_0^{\eps}(\rho^{\mE_n}_{R_nA_n}|R_n),
\end{equation}
where $\rho_{RA}^{\mE_n}=\Tr_{\sH_B^{\otimes n}}[\rho^{\mE_n}_{R_nA_nB_n}]$, with $\rho_{R_nA_nB_n}^{\mE_n}\in\states\left((\sH_R\otimes\sH_A\otimes \sH_B)^{\otimes n}\right)$ denoting the c-q extension of the state $\rho_{AB}^{\otimes n}$ corresponding to its pure-state ensemble decomposition $\mE_n$. By taking the appropriate limits ($n\to\infty$ followed by $\eps\to 0$) on either side of~(\ref{qwerty}) and employing Lemma~\ref{lemma:six} and Lemma~\ref{lemma:seven} below, one arrives at:
\begin{equation}\nonumber
 \begin{split} E_C(\rho_{AB})&\le\min_{\mE}\left\{-S_r(\rho_{RA}^\mE\|\rho_R^\mE\otimes\openone_A)\right\}\\
&\equiv \min_\mE H_r(\rho_{RA}^\mE|\rho_R^\mE)= E_F(\rho_{AB}).
\end{split}
\end{equation}
Inequality~(\ref{eq:to_prove}) is finally obtained by employing standard blocking arguments, see for example Ref.~\cite{ent_cost}. $\blacksquare$\bigskip

Before stating and proving Lemma~\ref{lemma:six} and Lemma~\ref{lemma:seven}, we need to recall some definitions and notations extensively used in the Quantum Information Spectrum Method~\cite{info-spect}. A fundamental quantity used in this approach is the \emph{quantum spectral inf-divergence rate}, defined as follows:
\begin{definition}[Spectral inf-divergence rate]
  Given a sequence of states $\hat\rho=\{\rho_n\}_{n=1}^\infty$, $\rho_n\in\states(\sH^{\otimes n})$, and a
  sequence of positive operators
  $\hat\sigma=\{\sigma_n\}_{n=1}^\infty$, where $\sigma_n$ acts on $\sH^{\otimes n}$, the \emph{quantum spectral
  inf-divergence rate} is defined in terms of the difference
  operators $\Delta_n(\gamma) = \rho_n - 2^{n\gamma}\sigma_n$ as
\begin{equation}
  \underline{D}(\hat\rho \| \hat\sigma) := \sup \left\{ \gamma :
    \liminf_{n\rightarrow \infty} \mathrm{Tr}\left[ \{ \Delta_n(\gamma)\ge 0\} \Delta_n(\gamma)\right] = 1 \right\}, \label{udiv}
\end{equation}
where the notation $\{X\ge 0\}$, for a self-adjoint operator $X$, is used to indicate the projector onto the subspace where $X\ge 0$.
\end{definition}

\noindent  We first note that, by definitions~(\ref{eq:19}) and~(\ref{eq:22}), we have:
\begin{equation}\label{zxcvbn}
\begin{split}
  &\min_{\mE_n}H_0^{\eps}(\rho^{\mE_n}_{R_nA_n}|R_n)\\
=&-\max_{\mE_n} \max_{\omega^n_{R_nA_n}\in \cqB(\rho_{R_nA_n}^{\mE_n})}\min_{\sigma^n_{R_n}}S_0(\omega^n_{R_nA_n}\|\sigma^n_{R_n}\otimes\openone_A^{\otimes n}).
\end{split}
\end{equation}
We then prove the following lemma:
\begin{lemma}\label{lemma:six}
For any bipartite state $\rho_{AB}$, with a pure-state ensemble decomposition
$\mE$, let $\rho_{RAB}^\mE$ denote its c-q extension. Then using the notation of (11), we have
\begin{align}
&\lim_{\eps\to 0}\lim_{n\to\infty}\left\{- \min_{\mE_n}H_0^{\eps}(\rho^{\mE_n}_{R_nA_n}|R_n) \right\}\nonumber\\
\ge&\max_{\mE}\min_{\hat\sigma_R}\underline{D}(\hat\rho_{RA}^{\mE}\|\hat\sigma_R\otimes\hat\openone_A),\label{eq:rhs}
\end{align}
where $\hat\rho_{RA}^{\mE}:=\left\{(\rho_{RA}^{\mE})^{\otimes n}\right\}_{n\ge 1}$, $\hat\openone_A:=\{\openone_A^{\otimes n}\}_{n\ge 1}$, and $\hat\sigma_R:=\{\sigma_R^n\in\states(\sH_R^{\otimes n})\}_{n\ge 1}$.
\end{lemma}

\noindent{\bf Proof}. Let $\bar\mE$ be the pure state ensemble decomposition of $\rho_{AB}$ for which the maximum on the r.h.s. of eq.~(\ref{eq:rhs}) is achieved, and let $\rho_{RA}^{\bar\mE}$ be its reduced state. Since $\bar\mE$ is fixed, in the following, we drop the superscript whenever no confusion arises, 
denoting  $\rho_{RA}^{\bar\mE}$ simply as $\rho_{RA}$.

Note that, for any fixed $\eps> 0$,
\begin{eqnarray}
&& - \min_{\mE_n}H_0^{\eps}(\rho^{\mE_n}_{R_nA_n}|R_n) \nonumber\\
&=&\max_{\mE_n}\max_{\omega_{R_nA_n}^n\in\cqB(\rho_{R_nA_n}^{\mE_n})}\min_{\sigma_{R_n}^n}S_0(\omega_{R_nA_n}^{\mE_n}\|\sigma_{R_n}^n\otimes\openone_A^{\otimes n})\nonumber\\
&\ge&\max_{\mE}\max_{\omega_{R_nA_n}^n\in\cqB((\rho_{RA}^{\mE})^{\otimes n})}\min_{\sigma_{R_n}^n}S_0(\omega_{R_nA_n}^n\|\sigma_{R_n}^n\otimes\openone_A^{\otimes n})\nonumber\\
&\ge&\max_{\omega_{R_nA_n}^n\in\cqB(\rho_{RA}^{\otimes n})}\min_{\sigma_{R_n}^n}S_0(\omega_{R_nA_n}^n\|\sigma_{R_n}^n\otimes\openone_A^{\otimes n}).\label{eq:here2}
\end{eqnarray}

For each $\sigma_{R_n}^n$ and any $\gamma\in\mathbb{R}$, define the projector \begin{equation}
  P_n^\gamma\equiv P_n^\gamma(\sigma_{R_n}^n):=\{\rho_{RA}^{\otimes n}- 2^{n\gamma}(\sigma_{R_n}^n\otimes\openone_A^{\otimes n}) \ge 0\}.
\end{equation}
Since the operator $\omega^n_{R_nA_n}$ in~(\ref{eq:here2}) is a c-q operator, it is clear that the minimization over $\sigma_{R_n}^n$ in~(\ref{eq:here2}) can be restricted to states diagonal in the basis chosen in representing c-q operators. Consequently, also $P_n^\gamma$ has the same c-q structure.

Next, let us denote by $\hat\rho_{RA}$ the i.i.d. sequence of states $\{\rho_{RA}^{\otimes n}\}_{n\ge 1}$. For any sequence $\hat\sigma_R:=\{\sigma_{R_n}^n\}_{n\ge 1}$, fix $\delta>0$ and choose $\gamma\equiv\gamma(\hat\sigma_R):= \underline{D}(\hat\rho_{RA}\|\hat\sigma_R\otimes\hat\openone_A) -\delta$. Then it follows from the definition~(\ref{udiv}) that, for $n$ large enough, \begin{equation}
  \Tr\left[P_n^\gamma\ \rho_{RA}^{\otimes n}\right]\ge 1-\eps^2/4,
\end{equation}
for any $\eps>0$. Further, define $\omega_{R_nA_n}^{n,\gamma}\equiv \omega_{R_nA_n}^{n,\gamma}(\sigma_{R_n}^n):=P_n^\gamma \rho_{RA}^{\otimes n} P_n^\gamma$, which is clearly in $B^{\eps}_{\mathrm{cq}}(\rho_{RA}^{\otimes n})$, due to the Gentle Measurement Lemma~\cite{gentle}.

Then, using the fact that $\Pi_{\omega_{R_nA_n}^{n,\gamma}} \le P_n^\gamma$, and Lemma~2 of~\cite{nila}, we have, for any fixed $\eps>0$,
\begin{eqnarray}
&& \lim_{n\to\infty}\frac 1n\,\{\textrm{r.h.s. of (\ref{eq:here2})}\}\nonumber\\
&\ge & \lim_{n\to\infty}\frac 1n\min_{\sigma_{R_n}^n} S_0(\omega^{n,\gamma}_{R_nA_n}\|\sigma_{R_n}^n\otimes\openone_A^{\otimes n})\nonumber\\
&= &\lim_{n\to\infty}\frac 1n\min_{\sigma_{R_n}^n}\left\{-\log\Tr\left[\Pi_{\omega^{n,\gamma}_{R_nA_n}}(\sigma_{R_n}^n\otimes\openone_A^{\otimes n})\right]\right\}\nonumber\\
 &\ge & \lim_{n\to\infty}\frac 1n\min_{\sigma_{R_n}^n}\left\{ -\log\Tr\left[P_n^\gamma(\sigma_{R_n}^n\otimes\openone_A^{\otimes n})\right]\right\}\nonumber\\
 &\ge &\min_{\hat\sigma_R} \gamma(\hat\sigma_R)\nonumber\\
&=& \min_{\hat\sigma_R}\underline{D}(\hat\rho_{RA}\|\hat\sigma_R\otimes\hat\openone_A)-\delta\nonumber\\
&=& \max_{\mE}\min_{\hat\sigma_R}\underline{D}(\hat\rho_{RA}^\mE\|\hat\sigma_R\otimes\hat\openone_A)-\delta
\end{eqnarray}
Since this holds for any arbitrary $\delta>0$, it yields the required inequality~(\ref{eq:rhs}) in the limit $\eps\to 0$. $\blacksquare$\bigskip

From (\ref{qwerty}), (\ref{zxcvbn}) and Lemma \ref{lemma:six} it follows that 
\begin{equation}
E_C(\rho_{AB}) \le - \max_{\mE}\min_{\hat\sigma_R}\underline{D}(\hat\rho_{RA}^\mE\|\hat\sigma_R\otimes\hat\openone_A),
\label{poi}
\end{equation}
with $\hat\rho_{RA}^\mE=\{(\rho_{RA}^\mE)^{\otimes n}\}_{n\ge 1}$.
Further, from the Generalized Stein's Lemma~\cite{brandao-plenio} and Lemma~4 in~\cite{q1}, the lemma below follows:
\begin{lemma}\label{lemma:seven}
For any given bipartite state $\rho_{RA}$,
\begin{equation}
\min_{\hat\sigma_R}\underline{D}(\hat\rho_{RA}\|\hat\sigma_R\otimes\hat\openone_A)=S_r(\rho_{RA}\|\rho_R\otimes\openone_A),
\end{equation}
where $\hat\rho_{RA}=\{\rho_{RA}^{\otimes n}\}_{n\ge 1}$, $\hat\sigma_R:=\{\sigma_{R_n}^n\in\states(\sH_R^{\otimes n})\}_{n\ge 1}$, and $\hat\openone_A:=\{\openone_A^{\otimes n}\}_{n\ge 1}$.
\end{lemma}

\noindent{\bf Proof}. Consider the family of sets $\mM:=\{\mM_n\}_{n\ge 1}$
\begin{equation}\label{eq:defsets}
  \mM_n:=\left\{\sigma_{R_n}^n\otimes\tau_{A_n}^n\in\states(\sH_R^{\otimes n}\otimes\sH_A^{\otimes n})\right\},
\end{equation}
such that $\tau_{A_n}^n:=(\openone_A/d_A)^{\otimes n}$. For this family, the Generalized Stein's Lemma~(Proposition III.1 of~\cite{brandao-plenio}) holds.

More precisely, for a given bipartite state $\rho_{RA}$, let us define
\begin{equation}
  S_\mM^\infty(\rho_{RA}):=\lim_{n\to\infty}\frac 1nS_{\mM_n}(\rho_{RA}^{\otimes n}),
\end{equation}
with $S_{\mM_n}(\rho_{RA}^{\otimes n}):=\min_{\omega^n_{R_nA_n}\in\mM_n}S_r(\rho^{\otimes n}_{RA}\|\omega^n_{R_nA_n})$, and
$\Delta_n(\gamma) = \rho_{RA}^{\otimes n} - 2^{n\gamma}\omega^n_{R_nA_n}$. From the Generalized Stein's Lemma~\cite{brandao-plenio} it follows that, for $\gamma>S^\infty_\mM(\rho_{RA})$, 
\begin{equation}
  \lim_{n\to\infty}\min_{\omega^n_{R_nA_n}\in\mM_n}\Tr\left[\{\Delta_n(\gamma)\ge 0\}\Delta_n(\gamma)\right]=0,
\end{equation}
implying that
$\min_{\hat\omega_{RA}\in\mM}\underline{D}(\hat\rho_{RA}\|\hat\omega_{RA})\le S^\infty_\mM(\rho_{RA})$.
On the other hand, for $\gamma<S^\infty_\mM(\rho_{RA})$,
\begin{equation}
  \label{eq:36}
  \lim_{n\to\infty}\min_{\omega^n_{R_nA_n}\in\mM_n}\Tr\left[\{\Delta_n(\gamma)\ge 0\}\Delta_n(\gamma)\right]=1,
\end{equation}
implying that $\min_{\hat\omega_{RA}\in\mM}\underline{D}(\hat\rho_{RA}\|\hat\omega_{RA})\ge S^\infty_\mM(\rho_{RA})$.
Hence $$\min_{\hat\omega_{RA}\in\mM}\underline{D}(\hat\rho_{RA}\|\hat\omega_{RA})= S^\infty_\mM(\rho_{RA}).$$

Finally, by noticing that, due to the definition~(\ref{eq:defsets}) of $\mM$,
\begin{equation}
\begin{split} \min_{\hat\omega_{RA}\in\mM}&\underline{D}(\hat\rho_{RA}\|\hat\omega_{RA})\\
=\min_{\hat\sigma_R}&\underline{D}(\hat\rho_{RA}\|\hat\sigma_R\otimes\hat\openone_A)+\log d_A,
\end{split}
\end{equation}
and that, due to Lemma~4 in~\cite{q1},
\begin{equation}
  S^\infty_\mM(\rho_{RA})=S_r(\rho_{RA}\|\rho_R\otimes\openone_A)+\log d_A,
\end{equation}
we obtain the statement of the lemma. $\blacksquare$\bigskip


\begin{thebibliography}{99}
\bibitem{footnote1} If $\Lambda$ is an LOCC transformation, $E_C^{(1)}(\rho_{AB};\eps)\ge E_C^{(1)}(\Lambda(\rho_{AB});\eps)$. To see this, note that, by Definition~\ref{def:ent-cost}, there exists an LOCC map $T$ that transforms $E_C^{(1)}(\rho_{AB};\eps)$ singlets into a state $\tilde\rho_{AB}$ which is $\eps$-close to $\rho_{AB}$. Then, by the monotonicity of the fidelity, $\Lambda(\tilde\rho_{AB})$ is also $\eps$-close to $\Lambda(\rho_{AB})$, meaning that the composed LOCC map $\Lambda\circ T$ transforms $E_C^{(1)}(\rho_{AB};\eps)$ singlets into a state which is $\eps$-close to $\Lambda(\rho_{AB})$. This implies that $ E_C^{(1)}(\Lambda(\rho_{AB});\eps)\le E_C^{(1)}(\rho_{AB};\eps)$.
\bibitem{footnote2} The presence of different smoothing parameters in the bounds of Theorem~\ref{thm_main} arises from having defined the dilution error in terms of the fidelity and the c-q smoothing in terms of the trace-norm. Such definitions have been adopted anyway, for sake of overall simplicity.
\end{thebibliography}
\end{document}